\documentclass{article}
\usepackage{jcappub}
\usepackage{tensor}
\usepackage[normalem]{ulem}

\usepackage{booktabs}
\usepackage{graphicx}
\usepackage{dcolumn}
\usepackage{bm}
\usepackage{amsmath,amssymb}
\usepackage{siunitx}
\usepackage{subfigure}
\usepackage{float}
\usepackage{caption}
\usepackage{rotating}
\usepackage{multirow}
\usepackage{xcolor}
\usepackage{ulem}

\title{Hubble parameter estimation via dark sirens with the LISA-Taiji network}
\author{Renjie Wang$^{1}$, Wen-Hong Ruan$^{2,3}$, Qing Yang$^{4}$, Zong-Kuan Guo$^{2,3,5}$, Rong-Gen Cai$^{2,3,5}$, Bin Hu$^{1,*}$}

\affiliation{$^{1}$Department of Astronomy, Beijing Normal University, Beijing, 100875, China}
\affiliation{$^{2}$CAS Key Laboratory of Theoretical Physics, Institute of Theoretical Physics, Chinese Academy of Sciences, P.O. Box 2735, Beijing 100190, China}
\affiliation{$^{3}$School of Physical Sciences, University of Chinese Academy of Sciences, No. 19A Yuquan Road, Beijing 100049, China}
\affiliation{$^{4}$College of Engineering Physics, Shenzhen Technology University, Shenzhen, 518118, China}
\affiliation{$^{5}$School of Fundamental Physics and Mathematical Sciences, Hangzhou Institute for Advanced Study, University of Chinese Academy of Sciences, Hangzhou 310024, China}

\emailAdd{bhu@bnu.edu.cn}

\abstract{
The Hubble parameter is one of the central parameters in modern cosmology, which describes the present expansion rate of the universe. 
Their values inferred from the late-time observations are systematically higher than those from the early-time measurements by about $10\%$. 
To come to a robust conclusion, independent probes with accuracy at percent levels are crucial. 
Gravitational waves from compact binary coalescence events can be formulated into the standard siren approach to provide an independent Hubble parameter measurement. 
The future space-borne gravitational wave observatory network, such as the LISA-Taiji network, will be able to measure the gravitational wave signals in the Millihertz bands with unprecedented accuracy. 
By including several statistical and instrumental noises, we show that within 5 years operation time, the LISA-Taiji network is able to constrain the Hubble parameter within $1\%$ accuracy, and possibly beats the scatters down to $0.5\%$ or even better. 
}


\begin{document}
\maketitle

\section{INTRODUCTION}

The measurement of the Hubble parameter has arrived at a crossroads~\cite{Freedman:2017yms}.
The values obtained from the early-time observables such as cosmic microwave background (CMB)~\cite{Aghanim:2018eyx} or big bang nucleosynthesis plus baryon acoustic oscillation~\cite{Abbott:2017smn} are indirect, because to get $H_0$ from those measurements one has to assume a cosmological model. 
Although these measurements are more precise compared with the late-time distance ladder~\cite{Riess:2019cxk,2019ApJ...882...34F}, in this way the resulted $H_0$ is cosmological model dependent. 
The distance ladder is a direct $H_0$ measurement. However, generally, it has more serious systematics, such as the reddening of the cepheid or red-giant branch stars, metallicity effects, etc~\cite{Riess:2019cxk,2019ApJ...882...34F}. 
Hence, the resulted values might be mis-calibrated due to the aforementioned astro-physical issues. 
A new independent $H_0$ measurement whose accuracy is better than $2\%$ is crucial in order to judge the current discrepancy~\cite{Verde:2019ivm,Wong:2019kwg}. 
Once this $2\%$ precision level is achieved, we shall give a priority to understand the systematics, especially the unknown ones, rather than simply to increase the sample volume. 

With the self-calibration by the theory of general relativity, gravitational waves (GWs) from compact binary coalescence (CBC) events open a completely novel observational window for $H_0$ determination~\cite{TheLIGOScientific:2017qsa,GBM:2017lvd,Guidorzi:2017ogy,Abbott:2017xzu,Abbott:2020khf}.
Depends on whether being associated with electromagnetic (EM) counterparts or not, GW events can be categorized into bright sirens~\cite{Schutz:1986gp,Holz:2005df} and dark sirens~\cite{Chen:2017rfc,Fishbach:2018gjp,Gray:2019ksv}.
The former demand fairly good synergies, which are extremely challenging for high redshift CBC events; while the latter, which do not rely on transient measurements, ask for a precise sky localisation to reduce the number of possible host galaxies.
Since the GW siren is a completely independent measurement, its result shall suffer from different systematics. 
Hence, it can shed some light on the Hubble tension.  Resolving this tension will bring us important implications. 
If the result from GW siren is consistent  with that from the early time measurements such as CMB, it would  imply that the current understandings of distance ladder systematics are not enough and the concordance model $\Lambda$CDM still works.   
On the other hand, if the result from GW siren agrees with  that from the distance ladders,  one needs to revise the $\Lambda$CDM model and there must exist some  new physics beyond the standard model of cosmology.
This is because several CMB experiments (including both space mission and ground-based telescopes), such as Planck~\cite{Aghanim:2018eyx}, SPT~\cite{Henning:2017nuy} and ACT~\cite{Choi:2020ccd}, are consistent with each others. 
Each of these experiments has special designs in itself. 
Hence, they shall have different systematics.

The Laser Interferometer Space Antenna (LISA)~\cite{Audley:2017drz}, a space-borne gravitational wave observatory, consists of three spacecrafts in an equilateral triangle configuration.
The separation distance between the spacecrafts is about $2.5$ million kilometres. 
The LISA constellation is in a heliocentric orbit behind the Earth by about $20^{\circ}$.
Taiji~\cite{Hu:2017mde} is a gravitational wave space facility proposed by the Chinese Academy of Sciences, with separation distance of $3$ million kilometres in a heliocentric orbit ahead of the Earth by about $20^{\circ}$.
The LISA-Taiji network~\cite{Audley:2017drz,Hu:2017mde}, will be able to localise the CBC events with unprecedented accuracy~\cite{Ruan:2019tje}.
As demonstrated previously, this advantage could help improve the Hubble constant determination.  

In this article we forecast the ability in estimating the  Hubble parameter  by using GW sirens data from the future space-based GW observatories.
Unlike stellar-mass binary black holes detected with aLIGO/Virgo~\cite{Abbott:2016blz}, for which the merger rate is observationally measured, there is no conclusive observational evidence for merging massive binary black holes (MBHs).
The models~\cite{Barausse:2012fy,Klein:2015hvg} adopted in this article are some  viable theoretical predictions up to our knowledge, and are also extensively studied in the literature.
The models are built by combining the cosmological galaxy formation history with the massive black hole binary (MBHB) formation dynamics. 
In details, the models follow the evolution of baryonic structures along a dark-matter merger tree according to the extended Press-Schechter formalism which is calibrated by N-body simulations.
Besides of the MBHs, the baryonic ingredients of the model include: the hot unprocessed inter-galactic medium, the cold metal-enriched inter-stellar medium, the stellar galactic disk, the stellar spheroid, the nuclear gas and the nuclear star cluster, etc. 
In the next section, we will highlight two of the most relevant aspects with GW emissions, namely, black hole seedings and time delays. 

\section{MODELS}

We consider $3$ different massive black hole formation models with different black hole seedings and time delays.
The ``light-seed'' scenario assumes that the black hole seeds are the remnants of population III stars (PopIII) with typical initial masses centered at $300M_{\odot}$, which is called  ``PopIII'' model.
In the ``heavy-seed'' scenario (assuming the critical Toomre parameter $Q_c=3$), MBHs arise from the collapse of protogalactic disks and already have the masses around $10^5M_{\odot}$ at high redshifts $z=15\sim20$.
Depending on whether there exist  the delays between MBHs and galaxy mergers or not,  these ``heavy-seed'' models  are named as ``Q3d'' and ``Q3nod'', respectively.

In ``popIII'' and ``Q3d'' models, after the dynamical friction phase, several hardening mechanisms are included.
In the gas-rich environments, the nuclear gas viscosity drags the merger of MBHB behind the merger galaxies.  
The typical delay is about $10\sim100$ Myr. 
In the gas-poor environments, three-body interactions with stars dominate the hardening process. 
It brings the MBHs together on a time scale about $5$~Gyr. 
If a MBHB stalls at about persec separation, a MBH triple system may be formed when a succeeding galaxy merger occurred. 
The typical delay is about $100$ Myr. 
This mechanism seems to work effectively only for the heavy systems with masses $>10^6\sim 10^7 M_{\odot}$; 
otherwise, the lightest MBH may also be ejected via the gravitational slingshot mechanism before the triple interactions trigger the merger of the inner binary. 
The details of the time delay prescriptions can be found in the reference \cite{Antonini:2015sza}.
One can view ``Q3d'' and ``Q3nod'' as the conservative and optimistic limits of the ``heavy-seed'' scenario.

For each of the $3$ models, we consider $2$ types of mission configurations  (``the LISA-Taiji network'', ``Taiji-only'') and $3$ different observation times ($1$-year, $3$-year, $5$-year).
And for each of combinations of the model, the mission configuration and the observation time, we generate $40$ sets of simulations including both the instrumental noise~\cite{Cornish:2018dyw,Guo:2018npi} and lensing noise~\cite{Hirata:2010ba,Bonvin:2005ps}.
Each set of simulations contains a few tens or a few hundreds CBC events according to different MBH formation models.
For each simulated CBC event, we estimate the posterior probability of the luminosity distance from the frequency-domain GW strains by using the Fisher information matrix method, which will be briefly mentioned in the following section. 

In order to determine $H_0$, we also need the redshift information from the host galaxy.
To do so, we sample galaxies uniformly in the comoving volume with the number density of $0.02$ Mpc$^{-3}$, according to the model~\cite{Barausse:2012fy}.
The adopted values of galaxy number density are located in the middle of the observational error bars (see Figure 1 of the reference\cite{Barausse:2012fy}). We verified that, within the observational uncertainty range $(2\times 10^{-3}, 6\times10^{-2})$, except for the blue events, the $H_0$ estimations from all the other types events\footnote{see the definition of different types  of events in the subsequent context.} are insensitive to the choice of the number density, due to the excellent sky localisation.
Then, we locate the possible host galaxies within $99\%$ ellipsoidal contours in the 3-dimensional parameter space spanned by the luminosity distance and observation solid angles.
For each of the host galaxy candidates, we assume their redshift uncertainties are negligible.
Finally, we present the Hubble parameter estimations based on these $720$ sets of simulations.
The flat $\Lambda$CDM model with $H_0=67.74$ and $\Omega_M=0.3$ is taken as our fiducial cosmological model.
The following results will not rely on the fiducial cosmological model significantly, especially for the local CBC events. It might be worth noting here that one should pay attention to the accuracy of the Hubble parameter $H_0$ through  our simulations, rather than the resulted $H_0$ value itself in this work. 

\section{RESULTS}

\begin{figure}
\centering
\subfigure{
\begin{minipage}[t]{0.33\linewidth}
\centering
\includegraphics[width=1.1\linewidth]{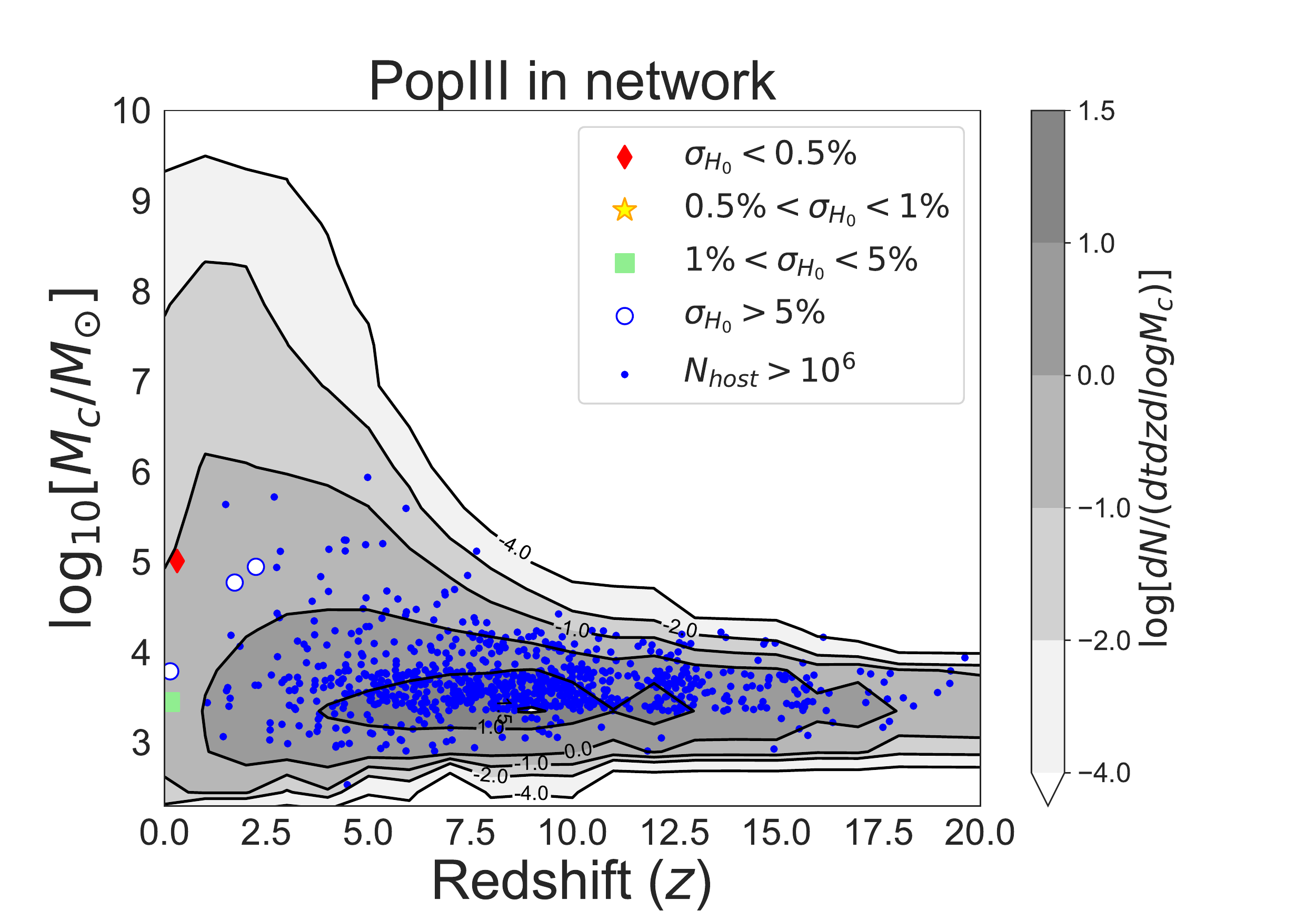}
\end{minipage}%
}%
\subfigure{
\begin{minipage}[t]{0.33\linewidth}
\centering
\includegraphics[width=1.1\linewidth]{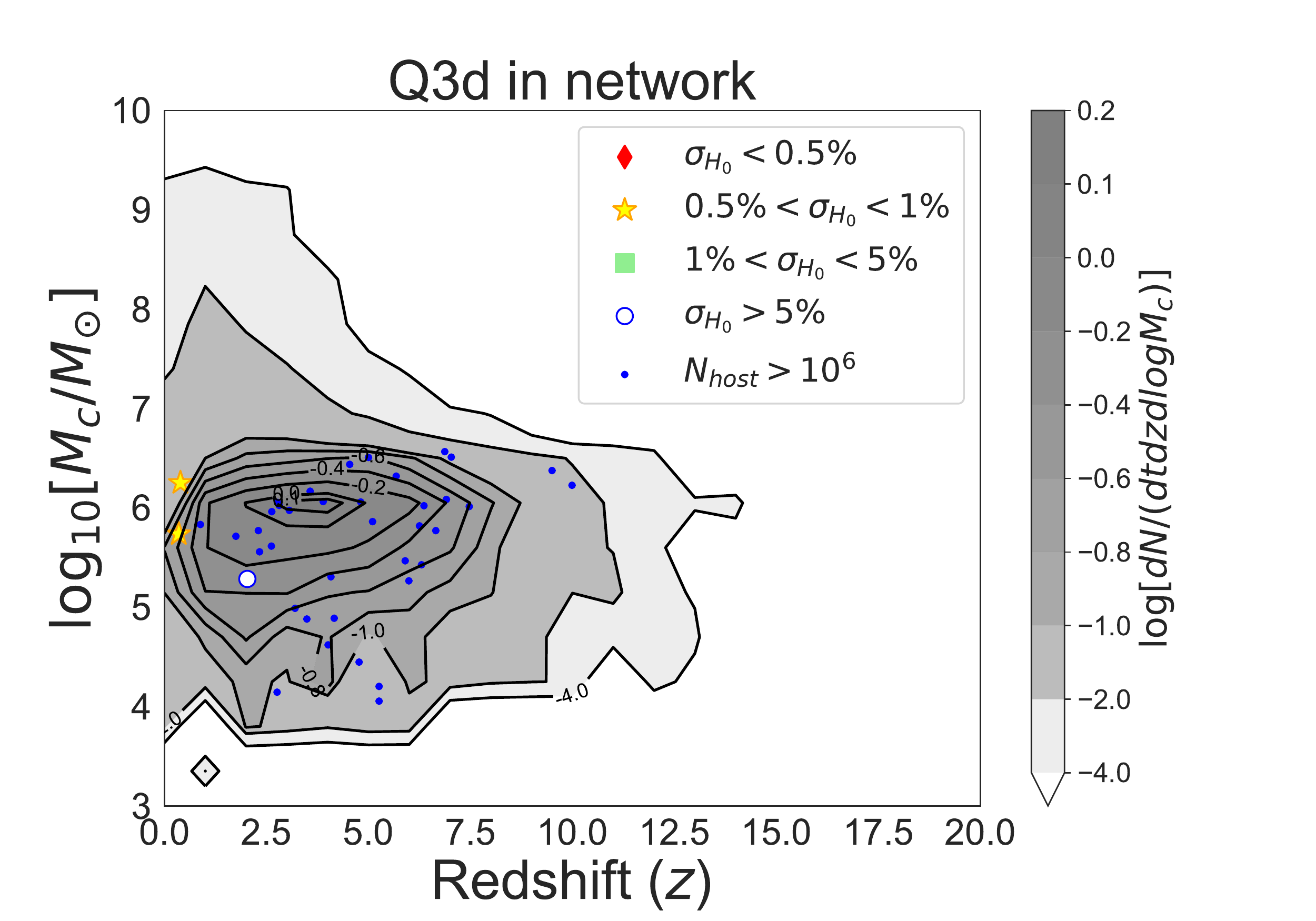}
\end{minipage}%
}%
\subfigure{
\begin{minipage}[t]{0.33\linewidth}
\centering
\includegraphics[width=1.1\linewidth]{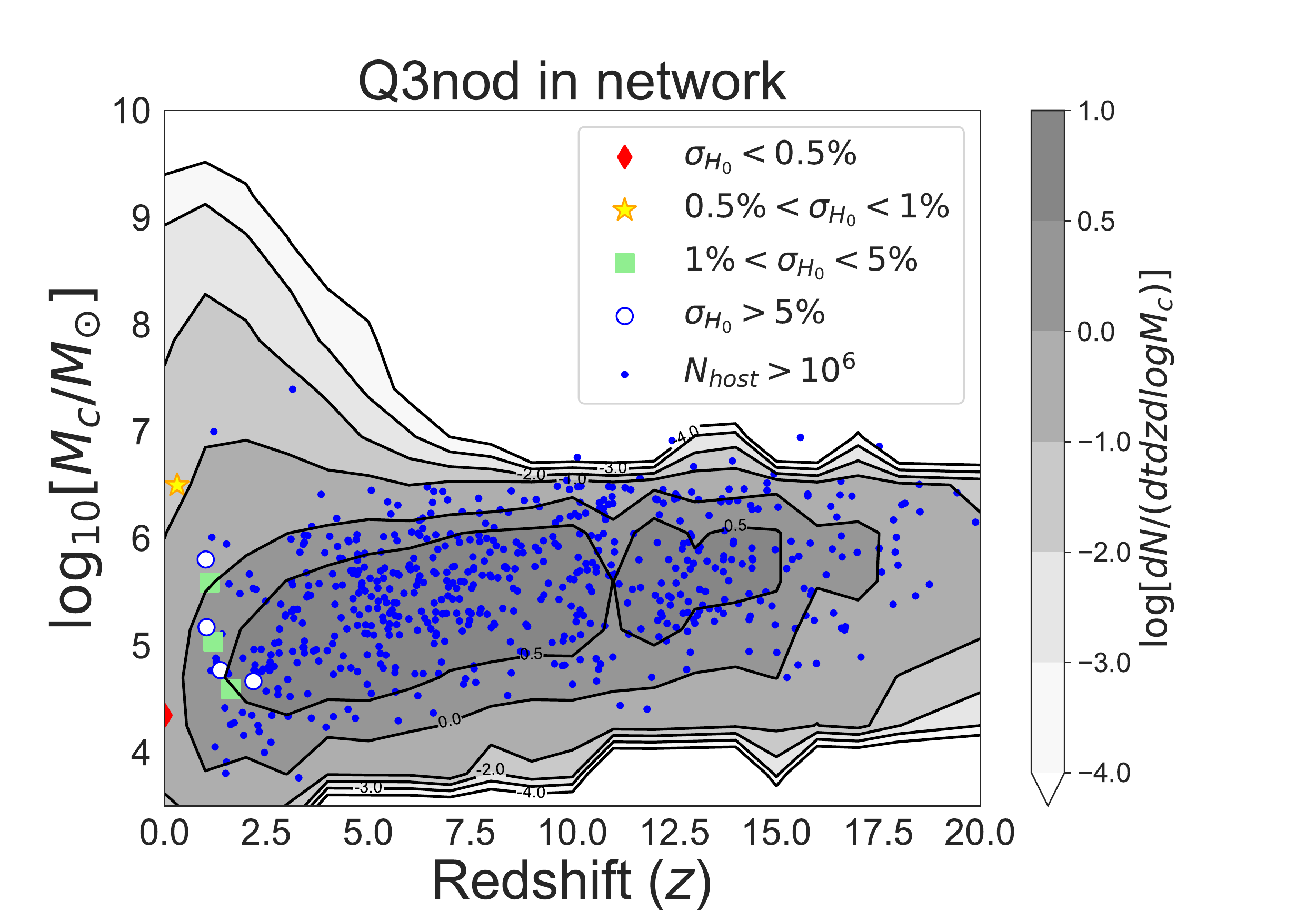}
\end{minipage}
}%
\newline
\centering
\subfigure{
\begin{minipage}[t]{0.33\linewidth}
\centering
\includegraphics[width=1.1\linewidth]{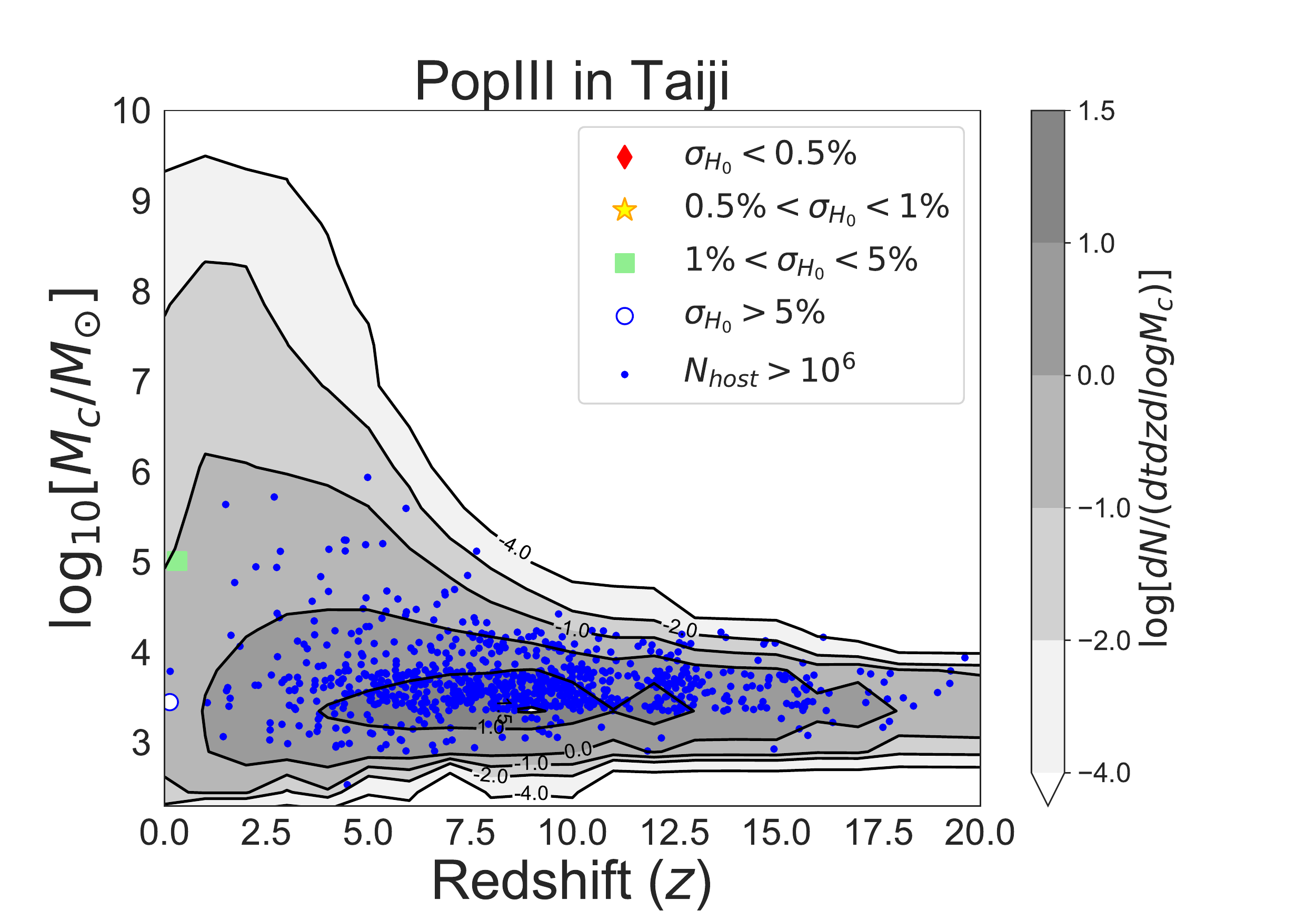}
\end{minipage}%
}%
\subfigure{
\begin{minipage}[t]{0.33\linewidth}
\centering
\includegraphics[width=1.1\linewidth]{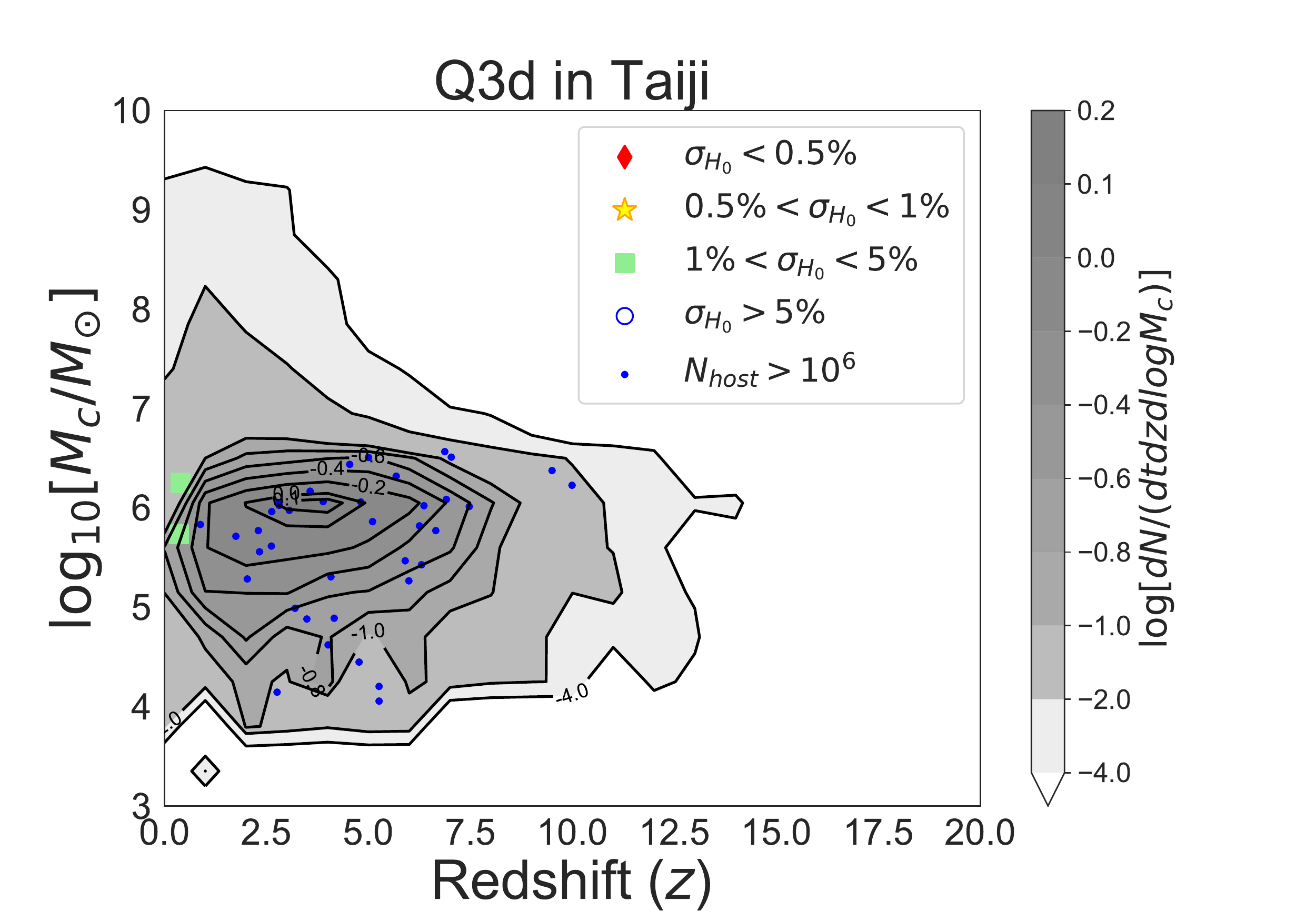}
\end{minipage}%
}%
\subfigure{
\begin{minipage}[t]{0.33\linewidth}
\centering
\includegraphics[width=1.1\linewidth]{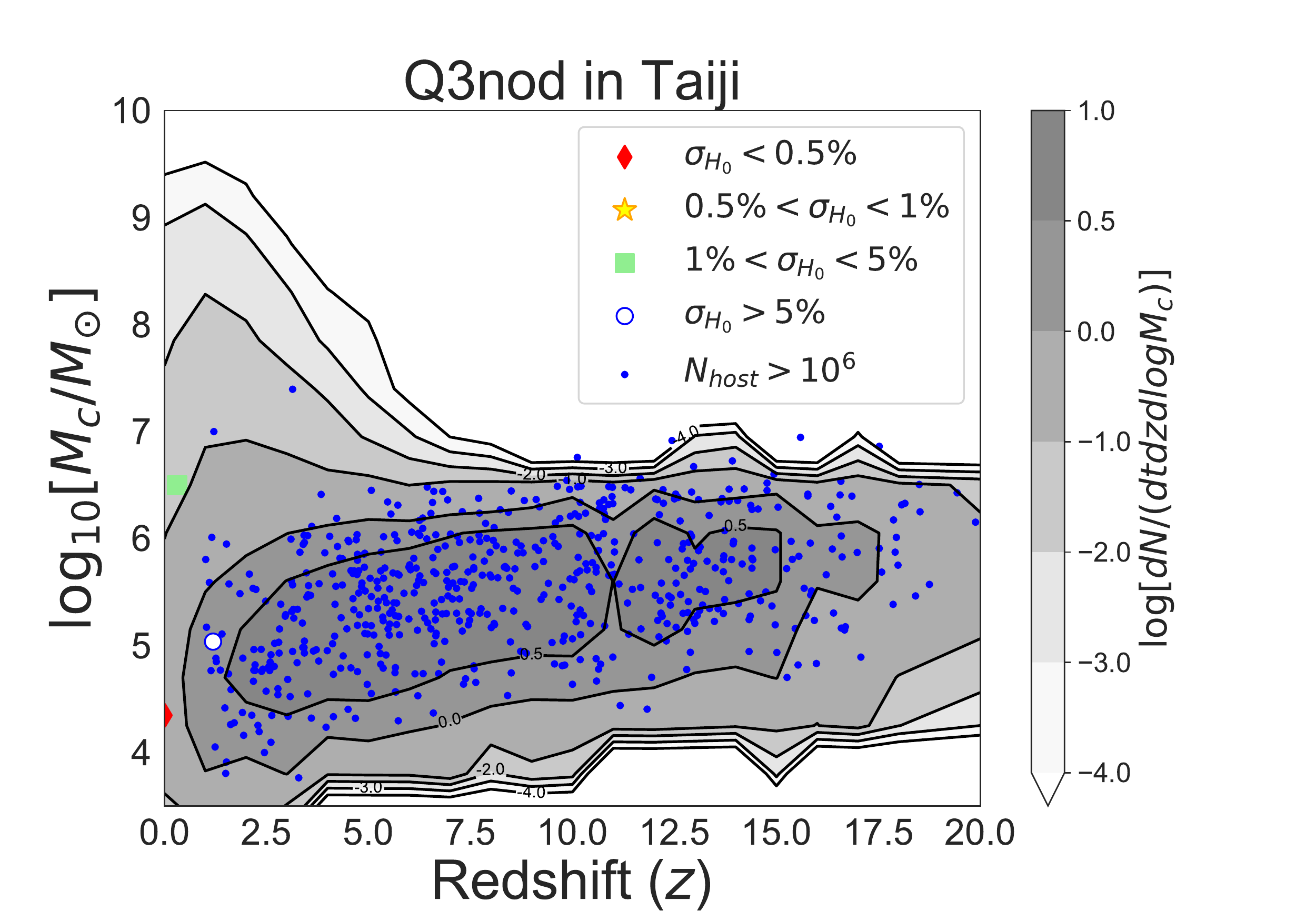}
\end{minipage}
}%
\centering
\caption{{\bf The simulated merger event rate distribution of massive black hole binaries (MBHBs) in redshift and chirp mass within 5-year observation time of the LISA-Taiji network and Taiji-only mission.}
Red diamonds ($\sigma_{H_0}/H_0<0.5\%$), yellow stars ($0.5\%-1\%$), green squares ($1\%-5\%$) as well as blue circles ($>5\%$) are the classified dark sirens according to their Hubble parameter
estimation accuracies. The filled blue spots are the unqualified dark sirens whose possible host galaxy numbers are more than $10^6$
due to the poor sky localisation. The background grey contours are the theoretical MBHB merger event rate distribution.
The first row are the results in the LISA-Taiji network, while the second row are for the  Taiji-only case.
The first, second and third columns are the predictions from $3$ different MBH models, namely PopIII, Q3d and Q3nod, respectively.
}
\label{fig:catalogs}
\end{figure}

\begin{figure}
\centering
\subfigure{
\begin{minipage}[t]{0.5\linewidth}
\centering
\includegraphics[width=1.1\linewidth]{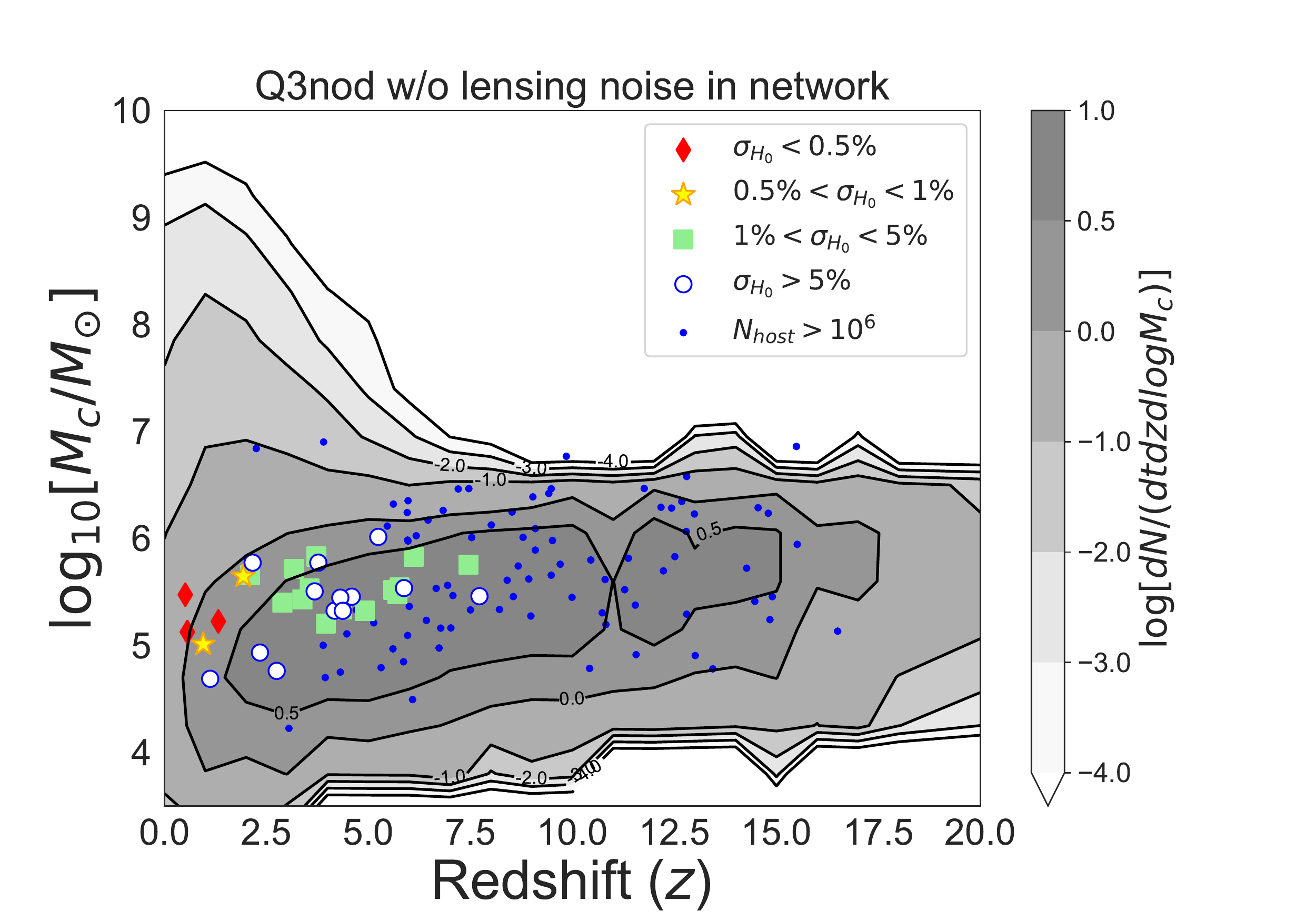}
\end{minipage}%
}%
\subfigure{
\begin{minipage}[t]{0.5\linewidth}
\centering
\includegraphics[width=1.1\linewidth]{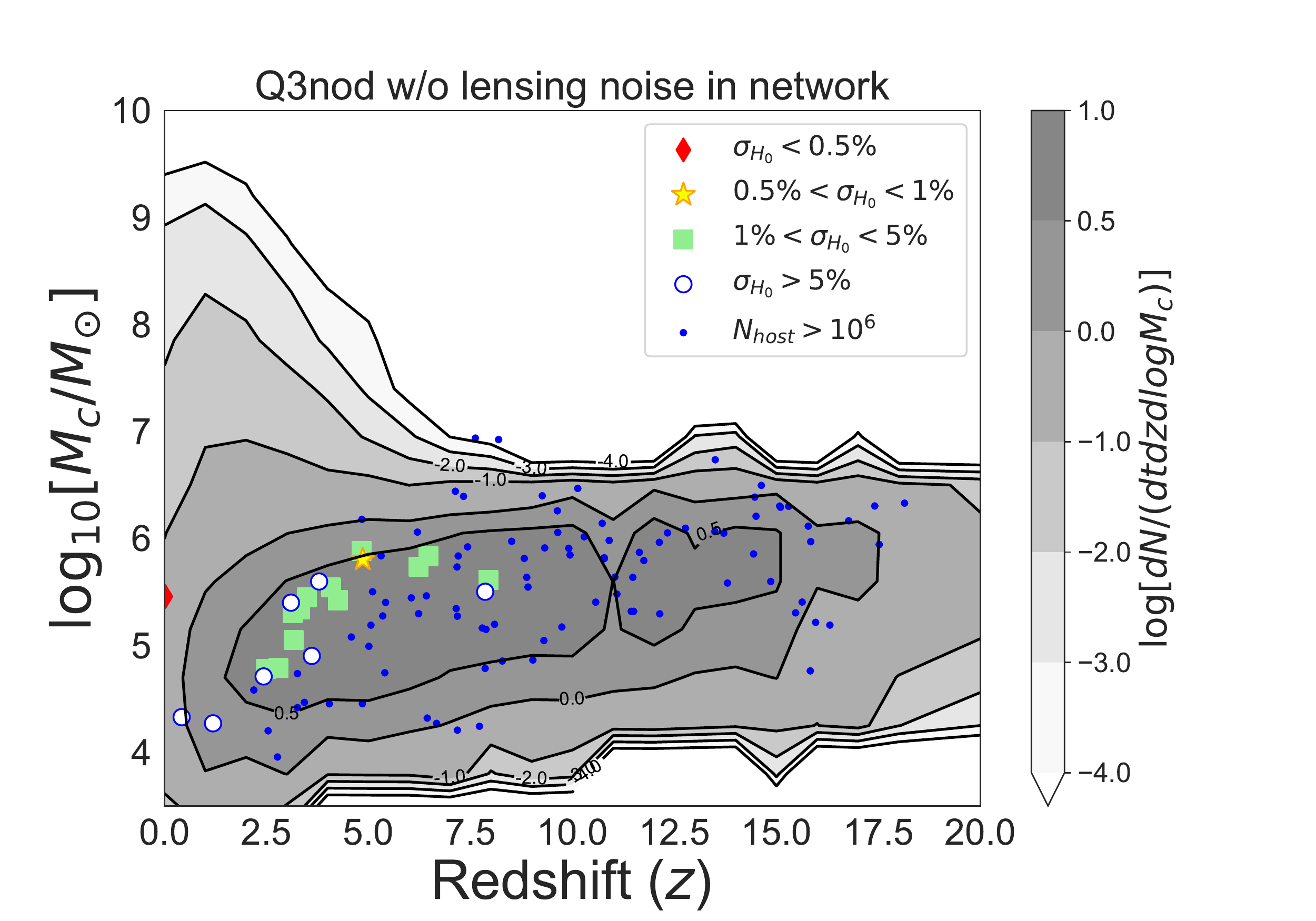}
\end{minipage}%
}%
\newline
\centering
\subfigure{
\begin{minipage}[t]{0.5\linewidth}
\centering
\includegraphics[width=1.1\linewidth]{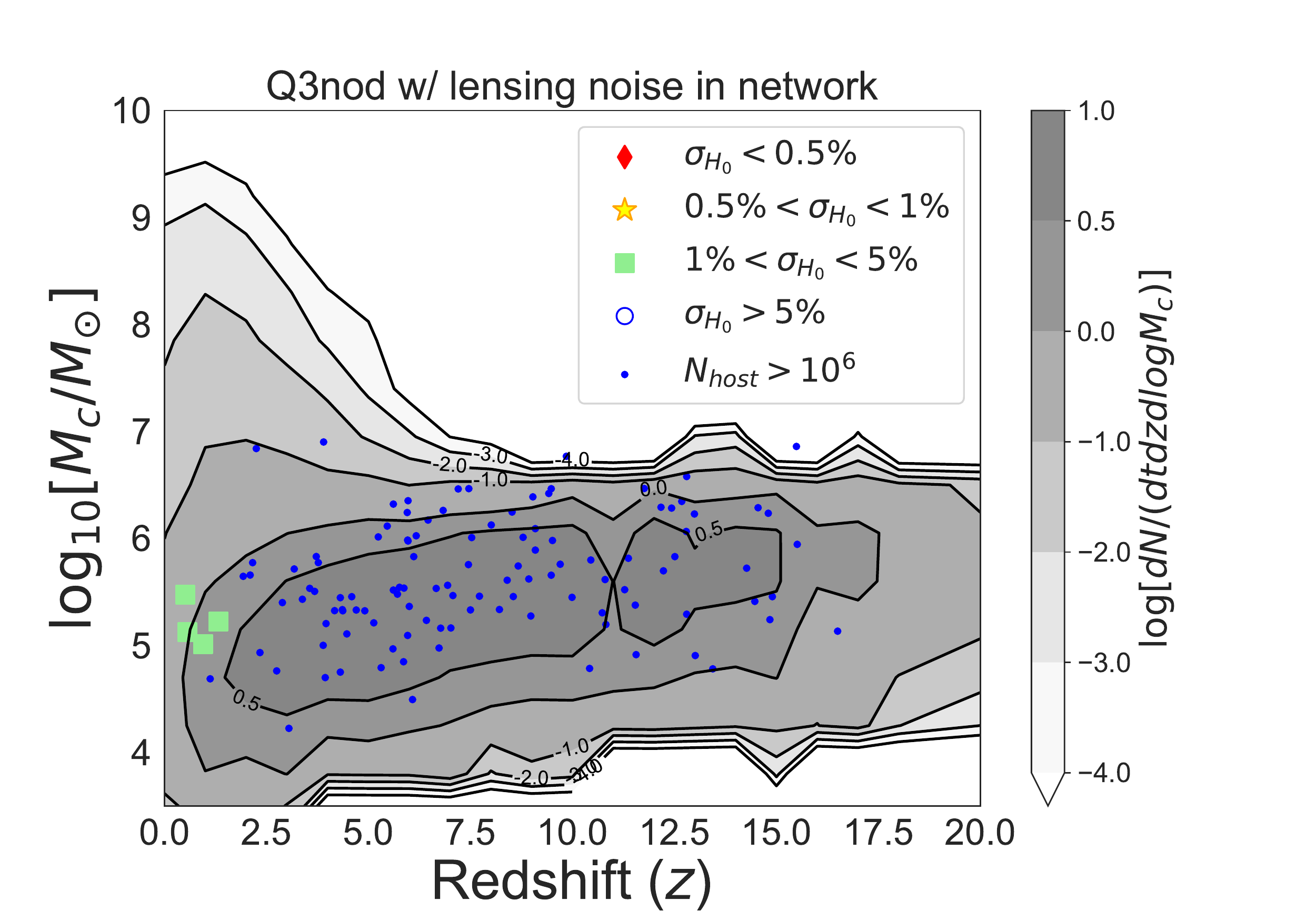}
\end{minipage}%
}%
\subfigure{
\begin{minipage}[t]{0.5\linewidth}
\centering
\includegraphics[width=1.1\linewidth]{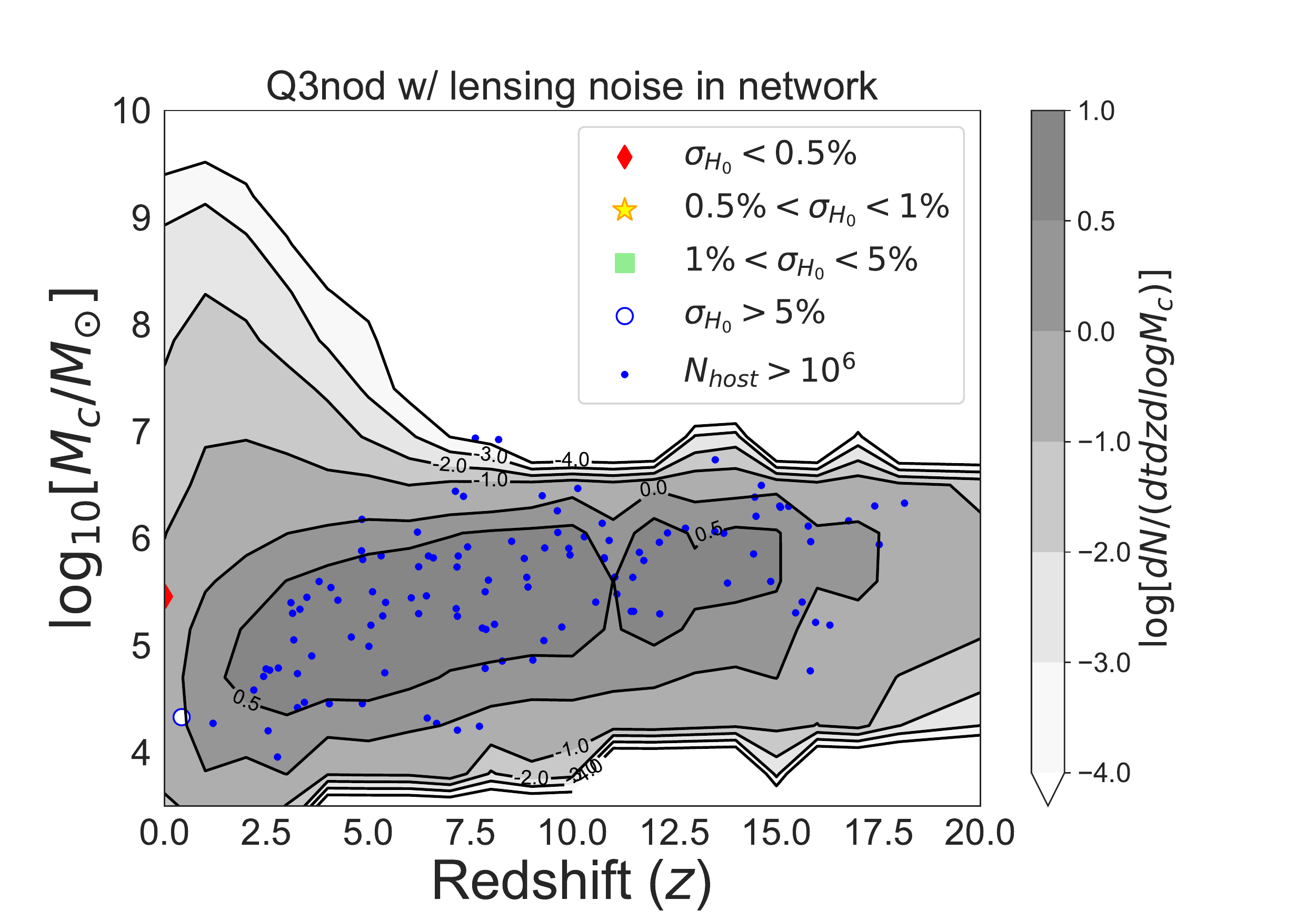}
\end{minipage}%
}%
\centering
\caption{{\bf Event distribution of dark sirens after 1-year observation of the LISA-Taiji network in the cases with/without lensing noise.}
We show $4$ mocked MBHB catalogs with $1$-year LISA-Taiji network observation.
The left and right are two individual realizations of Q3nod+network.
The top-left panel has a few diamond events with $z>0.5$; while in the top-right panel there is only one diamond at $z=0.02$.
The first row are the mocks without lensing noise. The second row are the mocks with lensing noise.
}
\label{fig:lens}
\end{figure}

In Figure~\ref{fig:catalogs}, we show one typical set of 5-year simulation in the LISA-Taiji network (the first row) as well as Taiji-only (the second row).
By the time of Taiji/LISA data collection, several $H_0$ measurements will hopefully achieve $1\%$ precision~\cite{Verde:2019ivm,Birrer:2020jyr}.
Hence, we classify all the qualified dark siren events into $4$ groups, namely diamond, gold, green and blue.
They correspond to the Hubble parameter with $<0.5\%$, $0.5\%-1\%$, $1\%-5\%$ and $>5\%$ accuracies, respectively.
Firstly, one can see that in all the $6$ panels, the number of qualified events is less than $10$.
This is because the nominal $H_0$ accuracies are extremely challenging.
Only the events, whose luminosity distance uncertainties are below percent levels, can qualify.
Secondly, all the qualified events are distributed below redshift $z=2.5$.
This is due to the lensing noise which will be demonstrated later.
Thirdly, the LISA-Taiji network can improve the results significantly, compared with the case of Taiji-only.
The upper and lower panels of the same columns are the results from the same CBC realizations.
Their differences lie in the mission configurations.
Taking the Q3d column as an example, the Taiji-only mission can capture $2$ green events after $5$-year observation.
In addition of capturing another blue event at redshift of $2$, the LISA-Taiji network is able to upgrade the $2$ green events in Taiji-only into the gold.
Last but not least, all diamond events are distributed in the very local universe.
This is also because, as long as $z>0.35$, the distance uncertainties induced by the unavoidable gravitational lensing do not meet the $H_0$ accuracy request.
In order to explain this more clearly, we show the event distribution in the cases with and without lensing noise in Figure~\ref{fig:lens}.
Two panels in the first row are those without lensing noise.
The left and right are two individual realizations of Q3nod+network.
The top-left panel has a few diamond and gold events in the redshift range $z>0.5$; while in the top-right panel there are one diamond event at $z=0.02$ and one gold event in the high-redshift ($z=4.86$).
One can see that without considering lensing noise the LISA-Taiji network could detect the qualified events all the way up to $z\simeq8$.
Two panels in the second row are those with lensing noise.
Comparing with the top-left, in the bottom-left panel all the original green and blue events fail the qualifications. Only the original $3$ diamond and $1$ gold events are survived, but downgraded into the greens.
However, the diamond in the top-right panel still keeps its identity in the bottom-right because lensing noise is negligible in the nearby universe.

\begin{figure}
\centering
\includegraphics[width=1.0\linewidth]{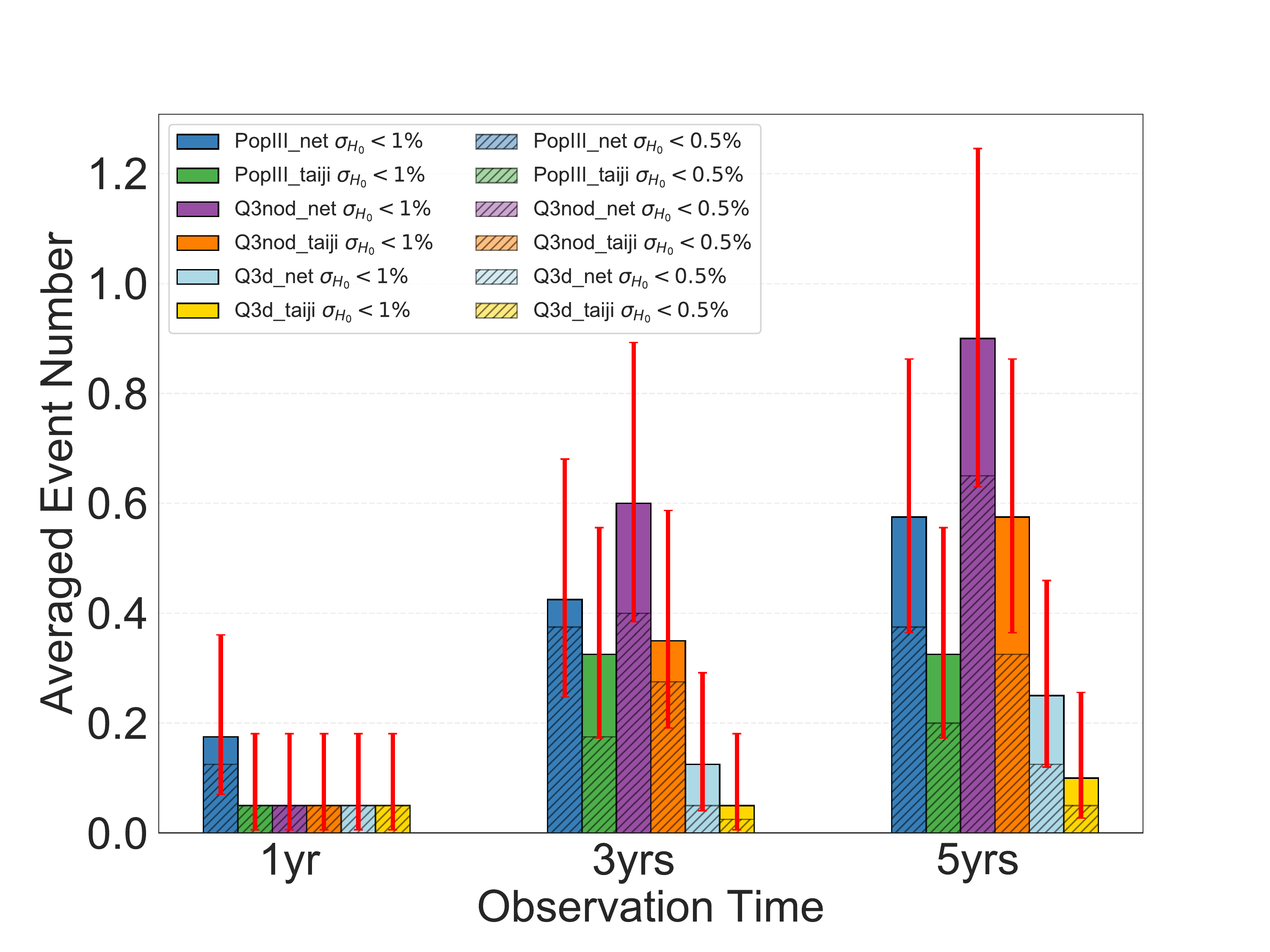}
\caption{{\bf Averaged event number in $1$-year, $3$-year and $5$-year observation time.}
Blue, green, purple, orange, cyan and yellow histograms denote the averaged event number in PopIII+network, PopIII+Taiji, Q3nod+network, Q3nod+Taiji, Q3d+network and Q3d+Taiji, respectively.
The unshaded histograms denote the dark sirens with $H_0$ accuracies better than $1\%$, namely diamond+gold events.
The red error bars denote  the $95\%$ confidence interval by assuming a Poisson distribution.
These error bars merely account for the statistical errors.
The shaded histograms denote  the dark sirens with $H_0$ accuracies better than $0.5\%$, namely diamond-only.
}
\label{fig:event_num}
\end{figure}

\begin{figure}
\centering
\includegraphics[width=1.0\linewidth]{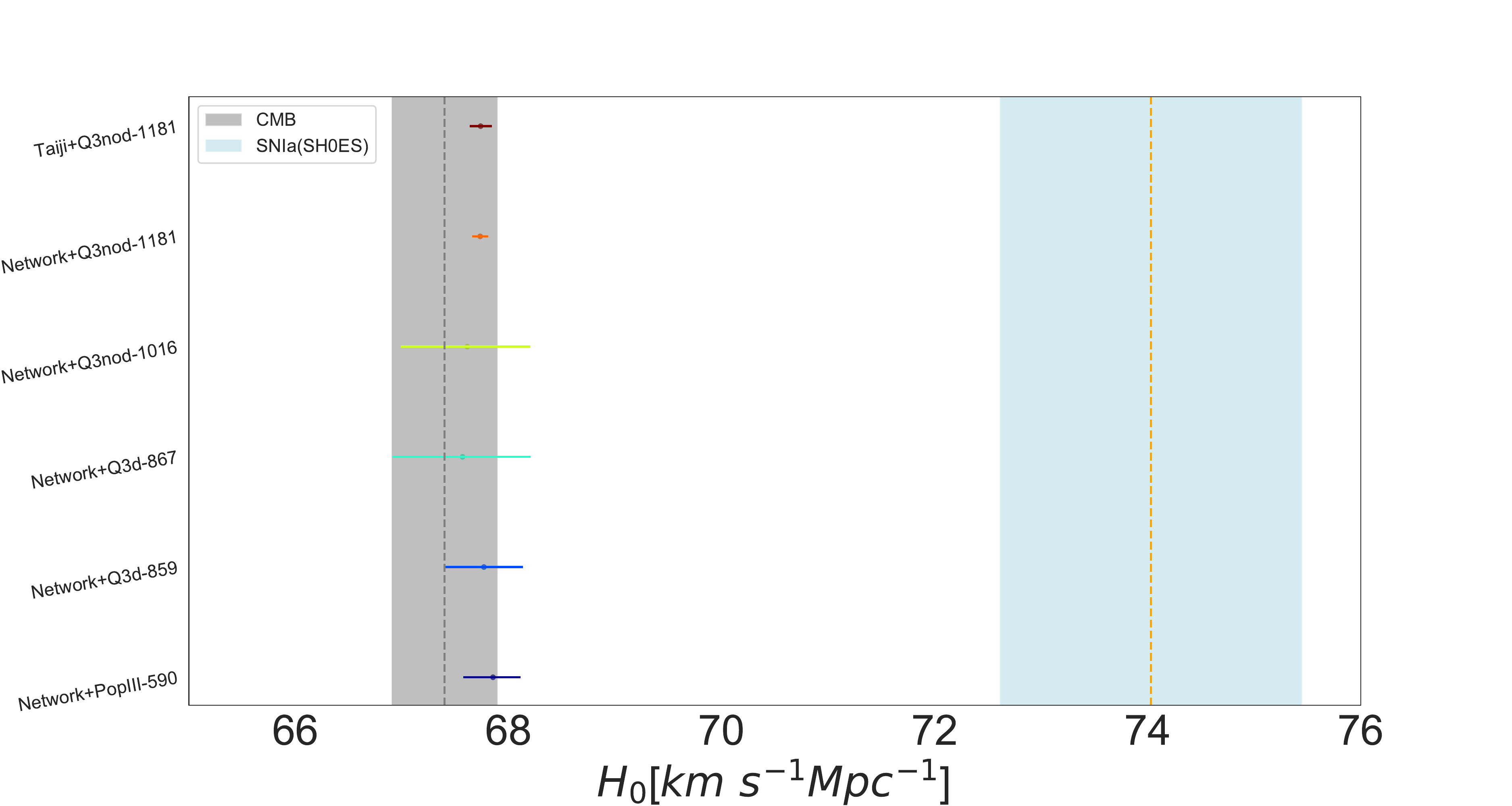}
\caption{{\bf Error estimation of the Hubble parameter from the diamond and gold events in the LISA-Taiji network and Taiji-only after $5$-year observation time.} The vertical grey ($H_0=67.4\pm0.5$ km s$^{-1}$ Mpc$^{-1}$) and cyan ($H_0=74.03\pm1.42$ km s$^{-1}$ Mpc$^{-1}$) bands denote  the present $H_0$ results from cosmic microwave background (Planck~\cite{Aghanim:2018eyx}) and SNIa (SH0ES~\cite{Riess:2019cxk}), respectively.
The fiducial value of the Hubble parameter is $H_0=67.74$.
The vertical axes are labelled as ``mission+event ID''.
Among these $6$ events, Net+Q3nod-1181, Taiji+Q3nod-1181 and Net+PopIII-590 are the diamond events.
The rests are the gold events.
}
\label{fig:best}
\end{figure}

In Figure~\ref{fig:event_num}, we show the averaged event numbers for the PopIII, Q3nod as well as Q3d models in $1$-year, $3$-year and $5$-year observation times, respectively.
In order to suppress the statistical errors, we compute each of the average numbers over $40$ sets of simulations.
From the statistics of $1$-year and $3$-year, we can not guarantee capturing $1$ diamond or gold event with $95\%$ confidence level.
After $5$-year network observation, for the Q3nod model the averaged event number with $H_0$ accuracy better than $1\%$ could reach $0.9$ and its $95\%$ confidence interval will up-cross unity.
We will very probably capture $1$ gold or diamond event after $5$-year network observation.
Comparing the shaded histogram (only diamond) with the unshaded one (diamond+gold) of Figure~\ref{fig:event_num}, we can see that the possibility of capturing a diamond event actually is higher than those of a gold event.
Again, this is still because of lensing noise.
For the PopIII model, the averaged event number accumulated in the network after $5$-year is about $0.58$, with the $95\%$ confidence interval in $0.36-0.86$.
For the Q3d model,  the averaged event number after $5$-year monitoring by the network is $0.25^{+0.20}_{-0.13}$.
The corresponding $5$-year numbers in the Taiji-only mission for both the PopIII and Q3nod models are about $2/3$ of those in the LISA-Taiji network.
And the Q3d events number in $5$-year Taiji-only mission is about half of the LISA-Taiji network case.
It implies that with the Taiji-only mission we are lack of confidence of capturing at least one diamond or gold event during $5$-year observation.
For the green ($1\%<\sigma_{H_0}<5\%$) events, the averaged numbers after $5$-year network observation are $4.05,0.88,0.38$ in the Q3nod, PopIII and Q3d models. respectively.
For the blue ( $\sigma_{H_0}>5\%$) events accumulated in the $5$-year network observation, the numbers are $8.00,3.60,1.23$.
The corresponding numbers of green (blue) events for $5$-year Taiji-only mission are  $0.58,0.23,0.25$ ($1.73,0.23,0.20$), respectively, for three models.
For elaborated statistics, we refer to Table 3 in the Supplements.

In Figure~\ref{fig:best}, we show the detailed $H_0$ results from diamond and gold events within the $5$-year observation, which have already been shown in Figure~\ref{fig:catalogs}.
The event Q3nod-1181 can qualify as the diamond in both the LISA-Taiji network and Taiji-only.
The former gives $H_0=67.73^{+0.08}_{-0.08}$, while the latter $H_0=67.74^{+0.10}_{-0.10}$.
They are both $0.1\%$ measurements.
This is because Q3nod-1181 is located only at $z=0.01$.
Both the LISA-Taiji network and Taiji-only are able to detect it with extremely high signal-to-noise ratio (${\rm SNR}\sim10^5$).
To ensure that such local event is not due to statistical fluke, we checked the redshift distribution of diamond events over $40$ sets of simulations under the ``Q3nod+network+5yrs'' configuration. 
We found that there are $5$ out of $26$ diamond events whose redshift equals to $0.01$. Besides, there are another $3$ diamond events whose redshifts are below $0.03$. Such local diamond events are typical in the ``Q3nod'' model. 
As for the event PopIII-590, the LISA-Taiji network can detect it as a diamond event ($H_0=67.85^{+0.26}_{-0.28}$, $0.4\%$ accuracy) with ${\rm SNR}\sim895$.
However, the Taiji mission can merely detect it as a green event ($H_0=67.81^{+1.08}_{-1.02}$, $1.6\%$ accuracy) with much lower SNR ($\sim564$).
Moreover, we can also tell the differences between two mission configurations by the sky area and numbers of possible host galaxies.
For PopIII-590, these two numbers in the LISA-Taiji network are $0.004~{\rm deg}^2$ and $3$ galaxies; while in the Taiji-only mission, they are $0.5~{\rm deg}^2$ and $1022$ galaxies.
From this example, we can clearly see that, the network can not only double the SNR, but also improve the sky localisation (reduce the numbers of possible host galaxies) significantly.
All these two aspects could help the measurement of the Hubble parameter by using dark sirens.
Besides of the diamond events, there are another $3$ gold events, namely Q3nod-1016, Q3d-867 as well as Q3d-859, which  could only be observed by the LISA-Taiji network.
Furthermore, there are $3$ green and $4$ blue events in the $5$-year network observation (see in the top-right panel of Figure~\ref{fig:catalogs}).
For detailed statistics, we refer to Table S1 and S2.
Although the green and blue events are not our major concerns, by combining these classified events, we can further reduce the $H_0$ error bars by at least $20\%$ ($H_0=66.61^{+1.80}_{-2.28}$ for joint-green, $H_0=65.41^{+2.70}_{-3.60}$ for joint-blue) w.r.t. the best individual cases in each categories  ($H_0=67.08^{+2.28}_{-2.46}$ for the best green, $H_0=66.31^{+3.42}_{-4.05}$ for the best blue). These can be seen in Figure S2.

\section{DISCUSSIONS}

GW siren is an independent $H_0$ measurement procedure.  Through the GW waveform, one is able to determine the luminosity distances to the GW sources.  Once the redshifts of GW sources 
are known through the bright sirens or dark sirens, one can obtain a relation between distance and redshift, through which $H_0$ is inferred. 
It {\it does not} mean that all the inferred $H_0$ values are cosmological model (eg. $\Lambda$CDM) independent.
In principle, if the Friedmann equation is used in the $H_0$ inference, the method is cosmological model  dependent; otherwise, it is not. 
One example of the model independent method is the SNIa distance ladder, in which the Hubble function or luminosity distance is Taylor expanded in terms of redshift. 
As shown in the reference~\cite{Riess:2016jrr}, the maximum redshift of this approach can be extended to $z_{\rm max}=0.4$.
Similar method can be applied to the GW sirens.  
In Table S1, we listed all the qualified dark sirens in Taiji. One can see that, $6$ out of $7$ events are distributed below redshift $0.4$. 
Moreover, in Table S2, all of the diamond and gold events in LISA-Taiji network are distributed below redshift $0.4$. 
These local events can be used to infer $H_0$ value via a cosmological model  independent method. 
However, there are some blue and green events from redshifts close to or higher than $1$. 
To utilise these data to infer $H_0$ value, one have to assume a background cosmological model. 
However, due to the poor quality of these data points, the resulted $H_0$ estimation from these events are the marginal results.

GWs cosmology, as a new exciting field, has a lot of unknowns in both theoretical modelings and observational systematics.
The results presented above are based on a simplified model setup.
There are lots of informative phenomena which we decide to turn a blind eye to.
First of all, we assume all MBHB mergers are dark.
As shown in our studies, the most important MBHB mergers for measuring $H_0$ are indeed those in the nearby universe.
For them, the EM counterpart observation may be possible~\cite{Tamanini:2016zlh}.
If EM counterparts can be identified, it will help to improve the sky localisation significantly.
Second, we do not consider the galaxy clustering effect.
The uniform distribution shall hold on average over sufficiently large volumes. However, in the small localisation ellipsoid, the clustering could help to reduce the $H_0$ error bars~\cite{Schutz:1986gp,MacLeod:2007jd,Chen:2017rfc,Gray:2019ksv,Mukherjee:2020hyn}. 
The clustering makes the redshift distribution more concentrated. Since the final $H_0$ posterior is the sum over all the possible redshifts, the narrower the redshifts are distributed, the faster the posterior will converge.   
In addition, although (both the bright and dark) sirens method asks for the redshift information, it does not ask for uniquely identifying the host galaxy, because the redshift is a smoothly varying quantity.  
Large scale structure predicts that fainter galaxies follow the clustering pattern of the more luminous galaxies. 
Hence, if the MBHB localisation ellipsoid is small enough, we may uniquely identify the central bright galaxy of the cluster where the true host faint galaxies reside in. 
In this case, we actually are able to upgrade the dark sirens into bright sirens.
Third, in order to avoid any theoretical bias, we do not utilise any other galaxy properties besides of redshift.
This is because our current understandings on the relationship between MBHs and dwarf galaxies are still unsatisfied.
If we could improve our knowledge on these aspects, we can aim at a particular type of galaxies instead of all the galaxies in the 3-dimensional contours.
As for the redshift uncertainties and the galaxy incompleteness, we have means to mitigate these problems.
Unlike the stellar binary black holes, MBHB populations are much less.
With the help of the space-based GW observatory network, we are able to localise each of them in a small area, such as $<10~{\rm arcmin}^2$.
Instead of using pre-existed galaxy catalogs, we could conduct deep optical and radio EM follow-ups for the limited diamond and gold events.
For (dwarf) galaxies with stellar masses $10^8M_{\odot}$ (corresponding to the central MBH with masses $10^5M_{\odot}$) at a luminosity distance of $1500$ Mpc, the K-band luminosity is about $24$ magnitude, which is completely visible for up-coming spectrograph observation, such as Thirty Meter Telescope~\cite{2014SPIE.9147E..24M}.
Based on these arguments, we believe we present an almost risk-free science case for the future space-borne GW mission.

\section{METHODS}

In this section, we present some essential aspects in the methodology of estimating $H_0$.

\subsection{\textbf{Fisher matrix}}

In order to simplify the calculation, we adopt the restricted post-Newtonian (PN) approximation of the GW waveform for the nonspinning MBHB~\cite{Sathyaprakash:2009xs}. For a nonspinning MBHB at a luminosity distance $d_L$, with component masses $m_1$ and $m_2$, total mass $M=m_1 + m_2$, symmetric mass ratio $\eta= m_1 m_2 / M^2$ and chirp mass $M_c= \eta^{3/5}M$, the frequency-domain version of the strain is given by~\cite{Ruan:2019tje,Ruan:2020smc}
\begin{equation}
\label{eq:strainf}
\tilde{h}(f)=-\left(\frac{5\pi}{24}\right)^{1/2}\left(\frac{GM_c}{c^3}\right)\left(\frac{GM_c}{c^2 D_{{\rm eff}}}\right)\left(\frac{GM_c}{c^3}\pi f\right)^{-7/6}e^{-i\Psi(f;M_c,\eta)},
\end{equation}
where $D_{\rm eff}$ is the effective luminosity distance to the source
\begin{equation}
\label{eq:deff}
D_{{\rm eff}}=d_L \left[F^{2}_{+}\left(\frac{1+{\rm cos}^2 \iota}{2}\right)^2+F^{2}_{\times} {\rm cos}^2 \iota
\right]^{-1/2},
\end{equation}
with the inclination angle $\iota$. The phase $\Psi$ depends on the coalescence time $t_c$ and the coalescence phase $\phi_c$~\cite{Allen:2005fk}. In this paper, $\Psi$ is calculated up to the second PN order. The response functions $F_{+}$, $F_{\times}$ depend on the sky direction of source $(\alpha, \delta)$ and the polarization angle $\psi$. For space-based GW detector such as LISA and Taiji, $F_{+}$ and $F_{\times}$ are functions of frequency~\cite{Ruan:2019tje}. In the calculation, the response functions of LISA and Taiji are obtained from the previous work~\cite{Rubbo:2003ap} with stationary phase approximation~\cite{Zhao:2017cbb}.

The Fisher matrix approach is employed in this paper to determine the uncertainty of parameter measurements for GW observation. For multiple detectors, the joint Fisher matrix is given by~\cite{Cutler:1997ta, Zhao:2017cbb}
\begin{equation}
\label{eq:FIM}
\Gamma_{ij}=\left(\frac{\partial_i \boldsymbol{d}(f)}{\partial \lambda_i}, \frac{\partial_j \boldsymbol{d}(f)}{\partial \lambda_j}\right),
\end{equation}
where $\boldsymbol{d}$ is written as
\begin{equation}
\label{ed:df}
\boldsymbol{d}(f)=\left[\frac{\tilde{h}_1 (f)}{\sqrt{S_1 (f)}},\frac{\tilde{h}_2 (f)}{\sqrt{S_2 (f)}},\cdots,\frac{\tilde{h}_N (f)}{\sqrt{S_N (f)}}\right]^{\rm T},
\end{equation}
and $\lambda_i$ denotes for the interested parameters. We consider the $9$ parameters of nonspinning MBHB ($M_c$, $\eta$, $d_L$, $\iota$, $\alpha$, $\delta$, $t_c$, $\phi_c$, $\psi$). Hence, $\Gamma$ is a $9$-dimensional matrix. Here, $S_i (f)$ is the noise power spectral density (PSD) of the $i{\rm th}$ detector and $\tilde{h}_i (f)$ is the frequency-domain GW strains. The noise-weighted inner product in Eq.~\eqref{eq:FIM} for two functions $a(t)$ and $b(t)$ is defined as
\begin{equation}
\label{eq:inprd}
(a, b)=2\int_{f_{\rm low}}^{f_{\rm up}}\left\{\tilde{a}(f)\tilde{b}^{*}(f)
+\tilde{a}^{*}(f)\tilde{b}(f)\right\}{\rm d}f\;.
\end{equation}
The upper cutoff frequency $f_{\rm up}$ is chosen as the innermost stable circular orbit (ISCO) frequency $f_{\rm isco}$ in the analysis, which is given by
\begin{equation}
\label{eq:fisco}
f_{\rm isco}=\frac{c^3}{6\sqrt{6}\pi GM}.
\end{equation}

Assuming the stationary Gaussian detector noise, the root-mean-square error of $\lambda_i$ is given by
\begin{equation}
\label{eq:rmse}
\sqrt{\langle \Delta \lambda_i^2 \rangle} =\sqrt{(\Gamma^{-1})_{ii}}.
\end{equation}
In our calculation, we use two Michelson-style data channels and the joint Fisher matrix is a sum of two Fisher matrices.

For a detected source at sky direction $(\alpha, \delta)$, the angular resolution is given by~\cite{Cutler:1997ta, Zhao:2017cbb}
\begin{equation}
\label{eq:domega}
\Delta \Omega_s=2\pi |{\rm sin}\alpha | \sqrt{\langle \Delta \alpha^2 \rangle \langle \Delta \delta^2 \rangle-\langle \Delta \alpha \Delta \delta \rangle^2},
\end{equation}
where $\langle \Delta \alpha^2 \rangle$, $\langle \Delta \delta^2 \rangle$ and $\langle \Delta \alpha \Delta \delta \rangle$ are given by the inverse of the Fisher information matrix. The uncertainty of $d_L$ can also be obtained according to Eq.~\eqref{eq:rmse}.

\subsection{\textbf{Lensing noise}}

\begin{figure}
\centering
\includegraphics[width=1.0\linewidth]{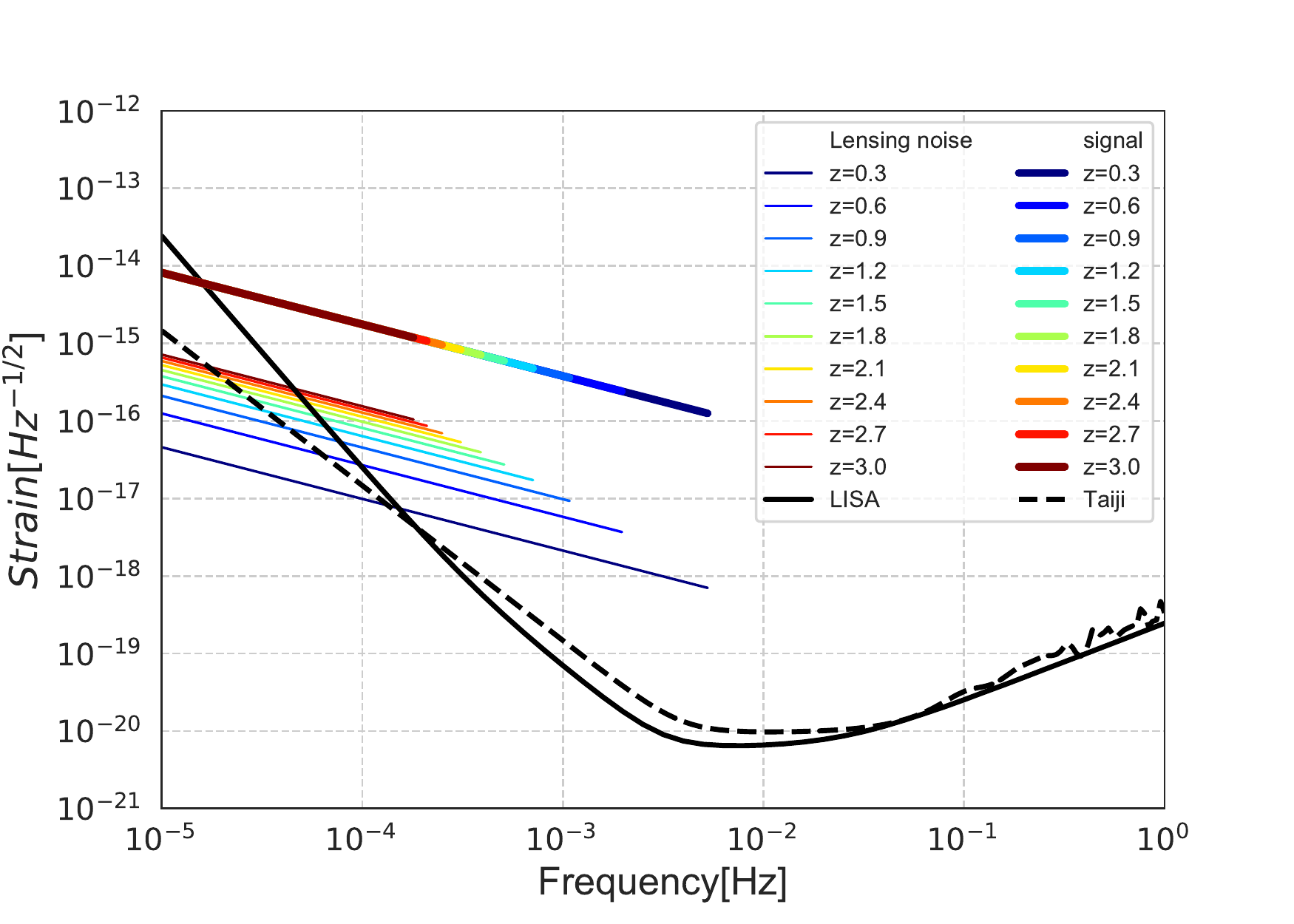}
\caption{{\bf Lensing and instrumental sensitivity curves in LISA and Taiji.}
The black solid and dashed curves are the instrumental sensitivity curves for LISA and Taiji, respectively.
The colored thin curves are the lensing noise of MBHB sources.
The colored thick curves are the GW signal strains.
Different colors denote different source redshifts.
}
\label{fig:sensi_curve}
\end{figure}

The effect of lensing magnification in GW observation is considered in the analysis.
In this paper, we model lensing effect via a stochastic noise in the luminosity distance.
The fitting formula of GW luminosity distance error due to lensing is given by~\cite{Hirata:2010ba}
\begin{equation}
\label{eq:sigmalens}
\sigma_{\rm lens}(z)=\frac{\Delta d_L}{d_L}=0.066\left[ \frac{1-(1+z)^{-0.25}}{0.25}\right]^{1.8}.
\end{equation}
 Hence, Eq.~\eqref{ed:df} can be rewritten as
\begin{equation}
\label{eq:dflens}
\boldsymbol{d}(f)=\left[\frac{\tilde{h}_1 (f)}{\sqrt{S_1 (f)+S_{1}^{\rm lens}(f)}},\frac{\tilde{h}_2 (f)}{\sqrt{S_2 (f)+S_{2}^{\rm lens}(f)}},\cdots,\frac{\tilde{h}_N (f)}{\sqrt{S_N (f)+S_{N}^{\rm lens}(f)}}\right]^{\rm T}\;,
\end{equation}
where the PSD of lensing noise for $i{\rm th}$ detector $S_{i}^{\rm lens}(f)$ is given by
\begin{equation}
S_{i}^{\rm lens}(f) = f \cdot \big|\tilde{h}_{i}^{\rm lens} (f) \big|^2,
\end{equation}
and $\tilde{h}_{i}^{\rm lens} (f)$ is obtained from
\begin{equation}
\tilde{h}_{i}^{\rm lens} (f) = \frac{1}{2}\left[\frac{\tilde{h}_i(f)}{1-\sigma_{\rm lens}(z)}-\frac{\tilde{h}_i(f)}{1+\sigma_{\rm lens}(z)}\right] \approx \sigma_{\rm lens}(z)\cdot \tilde{h}_i(f)\;.
\end{equation}
In Figure~\ref{fig:sensi_curve}, we show the lensing and instrumental sensitivity curves in LISA and Taiji space missions.
The black solid and dashed curves are the instrumental sensitivity curves for LISA and Taiji, respectively.
The colored thin curves are lensing noise of MBHB sources.
The colored thick curves are the GW signal strains.
Different colors stand for different source redshifts.
One can see that, from the redshift $0.3$ to $3$, lensing noises dominate over the instrumental noises in the frequency range from a few $10^{-5}$ Hz to a few mHz.
Lensing noise is the major component in the noise budget.
Here let us mention that because we want to demonstrate the relative lensing noise amplitude, we normalize all the primary GW strain signal from different redshifts, $\tilde{h}_i(f)$, with the same amplitude. 
That is the reason why all the signal curves align on the same line in Figure \ref{fig:sensi_curve}.

In this article, we simulate binary coalescence signals $\boldsymbol{d}(f)$ assuming a flat ${\rm \Lambda CDM}$ cosmology with $H_0=67.74$ and $\Omega_M=0.3$. The sky direction, inclination angle, coalescence phase and polarization angle are randomly chosen in the range of $\alpha \in [0, \pi]$, $\delta \in [0, 2\pi]$, $\iota \in [0, \pi]$, $\phi_c \in [0, 2\pi]$ and $\psi \in [0, 2\pi]$. The coalescence time of these samples are chosen to be $t_c = 0$ and $f_{\rm low}$ in~\eqref{eq:inprd} is randomly chosen between $10^{-5}$ Hz and the ISCO frequency. Moreover, we adopt the noise PSD without foreground confusion noise for LISA~\cite{Audley:2017drz, Cornish:2018dyw} and Taiji~\cite{Guo:2018npi}.
For the space-based GW mission, the confusion noise has three main components: short-period galactic binaries which are mostly from the white dwarf binaries (WDBs), short-period extragalactic binaries and compact objects (white dwarf, neutron star, stellar black hole) captured by MBHs~\cite{Barack:2004wc,Audley:2017drz}.  Among these components, the largest one is the galactic WDB background generated by millions of WDBs in the milky way. As shown in the Figure 1 of LISA white paper~\cite{Audley:2017drz}, in the frequency range $3\times10^{-4}-3\times10^{-3}$Hz, the galactic WDB background confusion noise could exceed the LISA instrumental noise, is about\footnote{The vertical axis of Figure 1 of LISA white paper~\cite{Audley:2017drz} is different from that in the Figure~\ref{fig:sensi_curve}. One has to divide the former with a factor $\sqrt{f}$ to convert it into the latter.} $2\times10^{-20}-6\times10^{-19}~{\rm Hz}^{-1/2}$.
However, from the Figure \ref{fig:sensi_curve}, one can see that the confusion noise level is about $3\sim4$ orders of magnitude smaller than the the targeted signals (the thick coloured curves is about a few $10^{-16}~{\rm Hz}^{-1/2}$).  
Hence, we argue that it shall be safe to neglect this component in the PSD.

\subsection{\textbf{Galaxy localisation}}

After generating GW signals, we need to firstly determine the CBC spatial localisation volumes based on the GW measurement uncertainties.
The simulated MBHB mergers are placed in the $3$-dimensional space spanned by the GW luminosity distance and sky direction angles, $(\log d_L$, $\alpha$, $\delta)$.
By marginalising over other $6$ model parameters, we get the $3$-dimensional covariance matrix, ${\bf Cov}\left[\log d_L, \alpha, \delta\right]$, of the source location parameters.
The probability density function of the source localisation can be written as
\begin{equation}
  f(\log(d_L), \alpha, \delta)=C \exp\left\{-\frac{1}{2}\Delta\theta^T {\bf Cov}[\log(d_L), \alpha, \delta]\Delta\theta\right\}\;.
\label{eq:gw_likelihood_den}
\end{equation}
Diagonalize the $3$-dimensional localisation covariance matrix~\cite{Yu:2020vyy, Amendola:2016wim}
\begin{equation}
 {\bf Cov}'(x,y,z)=({\bf v_1,v_2,v_3})^T{\bf Cov}[\log(d_L), \alpha, \delta]({\bf v_1,v_2,v_3})= \begin{pmatrix} \lambda_1&0&0\\0&\lambda_2&0\\0&0&\lambda_3 \end{pmatrix}\;,
\end{equation}
where $(\lambda_1, \lambda_2, \lambda_3)$ and $({\bf v_1, v_2, v_3})$ are the eigenvalues and eigenvectors of the original covariance ${\bf Cov}\left[\log(d_L), \alpha, \delta\right]$.
The orthogonal coordinates $(x, y, z)$ are linearly related with the original coordinates via the rotation
\begin{equation}
  \begin{pmatrix} x\\y\\z \end{pmatrix} =({\bf v_1,v_2,v_3})^T  \begin{pmatrix} \log(d_L)\\ \alpha \\ \delta \end{pmatrix}\;.
\end{equation}
With the orthogonal coordinates, the probability density function of the source location could be simplified
\begin{equation}
\label{eq:likehood}
  f(x,y,z)=C \exp\left\{-\frac{1}{2}\left[\frac{(x-\mu_x)^2}{\lambda_1}+\frac{(y-\mu_y)^2}{\lambda_2}+\frac{(z-\mu_z)^2}{\lambda_3}\right]\right\}\;,
\end{equation}
where ($\mu_x$, $\mu_y$, $\mu_z$) represent the coordinates of the simulated MBHBs and $C$ is a normalization factor.
This is a chi-square distribution with $3$ degrees of freedom.
Then we can draw an ellipsoid in $(x, y, z)$ space
\begin{equation}
  \frac{(x-\mu_x)^2}{\lambda_1}+\frac{(y-\mu_y)^2}{\lambda_2}+\frac{(z-\mu_z)^2}{\lambda_3}=\chi^2\;,
\end{equation}
with given confidence level which are characterised the value of $\chi^2$.
The volume enclosed by the ellipsoids are proportional to the CBC localisation probability.
In this work, we draw the ellipsoid with $99\%$ confidence level, which corresponds to $\chi^2=11.34$ according to the $3$-dimensional chi-square statistics.

\begin{figure}
\centering
\subfigure{
\begin{minipage}[t]{0.5\linewidth}
\centering
\includegraphics[width=1.0\linewidth]{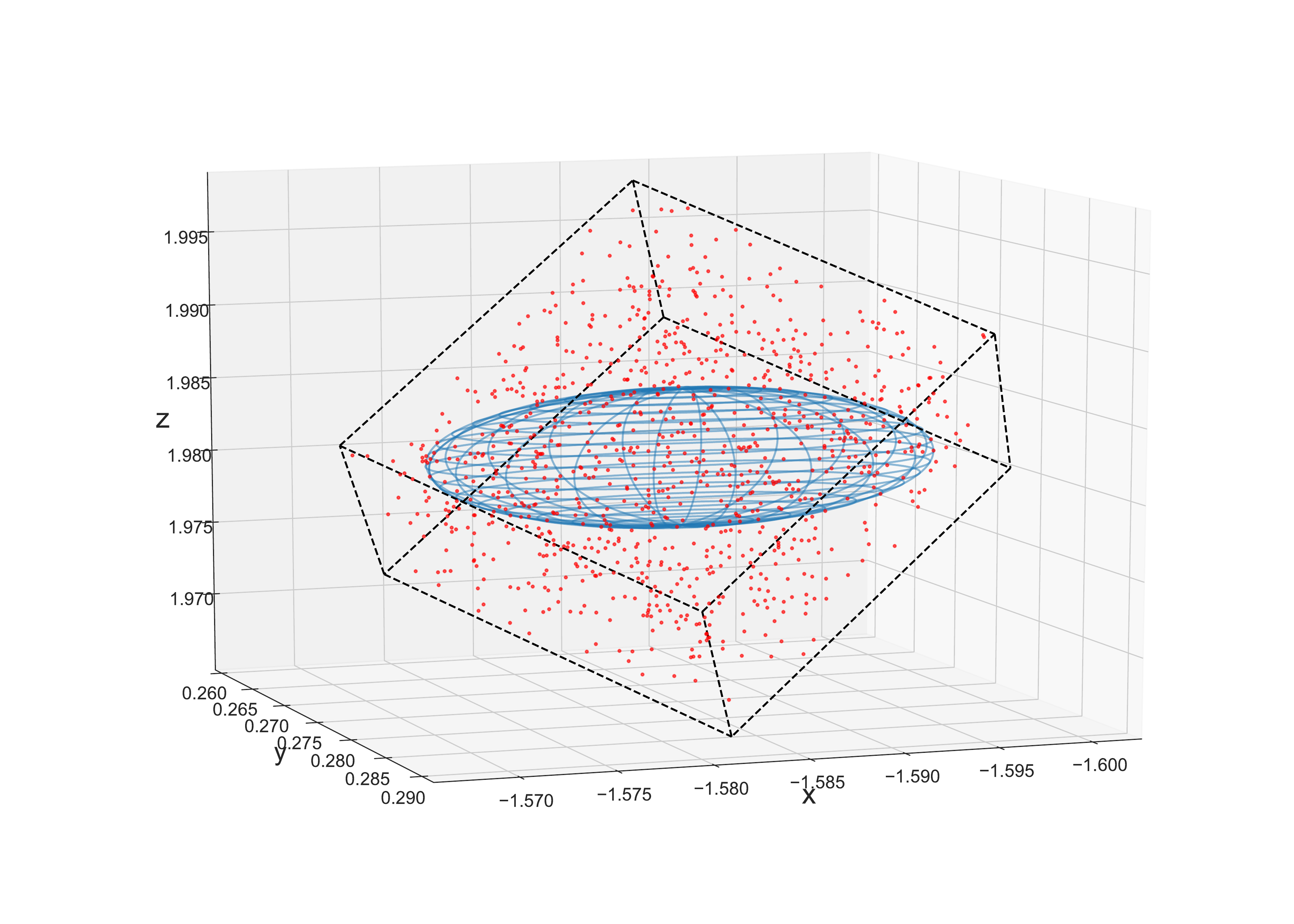}
\end{minipage}%
}%
\subfigure{
\begin{minipage}[t]{0.5\linewidth}
\centering
\includegraphics[width=1.0\linewidth]{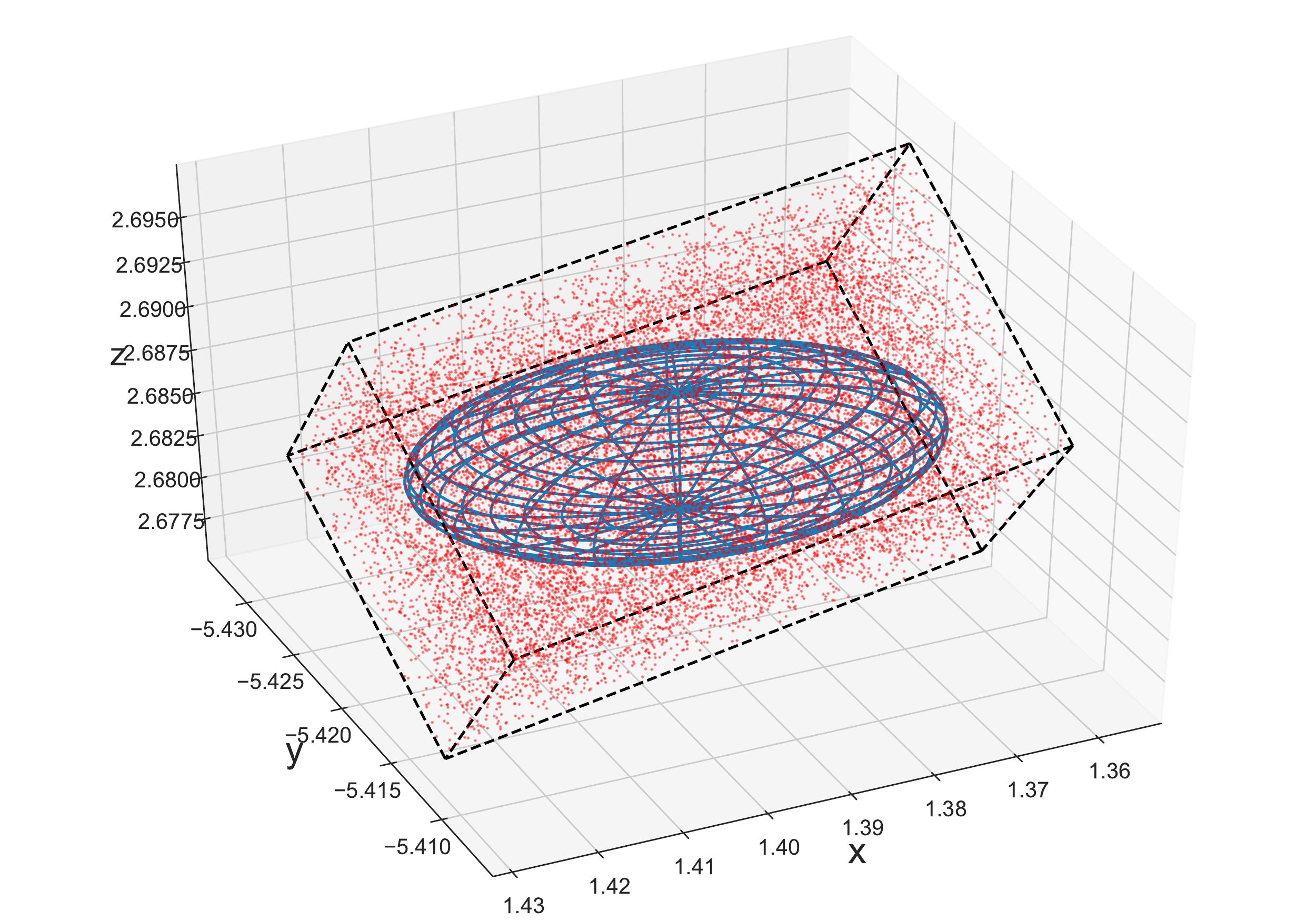}
\end{minipage}%
}%
\centering
\caption{{\bf Examples of CBC spatial localisations in the LISA-Taiji network.}
The blue nested ellipsoids are the $99\%$ confidence regime for CBC localisation.
The red points are the galaxy samplings.
The left and right ellipsoids enclose $185$ and $3505$ galaxies, respectively.
}
\label{fig:ellip}
\end{figure}

Then we populate the host galaxy candidates around the targeted ellipsoids.
To make sure the galaxy samplings can cover the targeted ellipsoids, we sample the galaxy in the $4\sigma$ ($99.99\%$) confidence regimes.
The galaxies are uniformly sampled in the comoving volume with the number density of $0.02$ Mpc$^{-3}$, according to the model~\cite{Barausse:2012fy}.
In Figure~\ref{fig:ellip}, we show $2$ examples of CBC spatial localisationf in the LISA-Taiji network.
The left and right ellipsoids enclose $185$ and $3505$ galaxies, respectively.
The background grey axies are the orthogonal coordinates $(x, y, z)$.
The foreground black frames are the original $(\log d_L$, $\alpha$, $\delta)$ coordinates.
The blue nested ellipsoids are the $99\%$ confidence regime for CBC localisation.
The red points are the galaxy samplings.
We assume all the galaxy redshifts can be measured with negligible errors. 
This is a reasonable assumption compared with the luminosity distance errors obtained by GW measurement.
The reasons are what follows.
For diamond events, due to the perfect sky localisation, we are able to conduct the spectroscopic follow-up. In this case, we shall safely neglect the redshift uncertainty. 
For the other type of events, once we consider the clustering effect, it will help the determination of the redshift. Instead of finding the correct host galaxy, we can search for the brightest central galaxy in the clusters where the true host resides in.
In this case, we can conduct photometric observation to the larger volume. As predicted for the Vera Rubin Observatory, previously referred to as the Large Synoptic Survey Telescope~\cite{Abell:2009aa,Padmanabhan:2020zxj}, 
in the redshift range $0 < z < 4$ the photometric redshift errors, $\sigma_z/(1+z)$, must be smaller than $0.05$, with a goal of $0.02$. The corresponding number for WFIRST (now renamed as Roman Space Telescope)~\cite{Hemmati:2019gyg} is about $0.002$.

\subsection{\textbf{Hubble parameter estimation}}
Finally, we come to estimate the posterior probability distribution of $H_0$ given both GW data ($d_{GW}$) and EM counterparts data ($d_{EM}$).
According to the Bayes theorem, the posterior of a single CBC event is
\begin{equation}
  p(H_0|d_{GW},d_{EM})=\frac{p(d_{GW},d_{EM}|H_0)p(H_0)}{\beta(H_0)},
\end{equation}
where $p(H_0)$ represents for the prior probability of $H_0$ and $\beta(H_0)$ for the evidence.
Since the two measurements are independent, we treat the joint GW and EM likehood, $p(d_{GW},d_{EM}|H_0)$, as the product of two individual likelihoods~\cite{Chen:2017rfc,Soares-Santos:2019irc}.
We marginalize over all the other variables except for the luminosity distance $d_L$, the solid angle $\hat{\Omega}_{\rm GW}$ of the GW source, the true host galaxy redshift $z_i$ and its solid angle $\Omega_i$.
Finally, the joint likelihood for $H_0$ can be written as
\begin{multline}
  p(d_{GW},d_{EM}|H_0)\propto \sum_i w_i\iiiint p(d_{GW}|d_L,\hat\Omega_{\rm GW})p(d_{EM}|z_i,\Omega_i)\delta(d_L-d_L(z_i,H_0)){} \\
  \times \delta(\hat\Omega_{\rm GW}-\Omega_i)p_0(z_i,\Omega_i)dd_Ld\hat \Omega_{\rm GW}dz_id\Omega_i\;,
\end{multline}
where $w_i$ are the weights for each individual galaxies.
Since we do not use other galaxy properties besides of their redshifts, we set the weighting factor equals to unity for all galaxies.
As mentioned before, we assumed galaxies are uniformly distributed in the comoving volume.
Hence, the prior, $p(z_i,\Omega_i)$, for galaxy redshift space distribution can be written as~\cite{Soares-Santos:2019irc}
\begin{equation}
  p_0(z_i,\Omega_i)\propto \frac{1}{V_{max}} \frac{d^2V}{dz_id\Omega} \propto \frac{1}{V_{max}} \frac{\chi^2(z_i)}{H(z)}\;,
\end{equation}
where $\chi(z)$ is the comoving distance to the galaxy.

Assuming we precisely know the galaxy redshift space position $(z_i,\Omega_i)$, we can express the EM counterparts likelihood as the products of delta functions
\begin{equation}
  p(d_{EM}|z_i,\Omega_i)\propto \prod \delta(z_{i,obs}-z_i)\delta(\Omega_{i,obs}-\Omega_i)\;.
\end{equation}
The GW likelihood, $p(d_{GW}|d_L,\hat\Omega_{\rm GW})$, can be calculated according to Eq.~\eqref{eq:gw_likelihood_den}.
Therefore, the final posterior for $H_0$ becomes~\cite{Yu:2020vyy}
\begin{equation}
  p(H_0|d_{GW},d_{EM})\propto \frac{p(H_0)}{\beta(H_0)} \sum p(d_{GW}|d_L(z_i,H_0),\Omega_i)p_0(z_i,\Omega_i)\;.
\end{equation}

\section{ACKNOWLEDGEMENTS}
We acknowledge Enrico Barausse and Hai-Bo Yuan for helpful discussions. BH and RJW are supported by the National Natural Science Foundation of China Grants No. 11690023, No. 11973016 and No. 11653003. RGC is supported by the National Natural Science Foundation of China Grants No.11690022, No.11821505, No. 11991052,
 No.11947302 and by the Strategic Priority Research Program of the Chinese Academy of Sciences Grant No. XDB23030100  and the Key Research Program of Frontier Sciences of CAS. ZKG and WHR are Supported by the National Natural Science Foundation of China Grants No.11690021, 12075297 and 11851302.

\bibliographystyle{JHEP}
\bibliography{ref}
\end{document}